\documentstyle[12pt,aasms4]{article}

\lefthead{F. J. Masci et al.}
\righthead{Soft X-ray Absorption in Parkes-Quasars}

\begin{document}

\title{Red Parkes-Quasars: Evidence for Soft X-ray Absorption}

\author{F. J. Masci\altaffilmark{1}}
\affil{Astrophysics Group, School of Physics,
University of Melbourne, Victoria 3052, Australia}

\author{M. J. Drinkwater\altaffilmark{2}}
\affil{Department of Astrophysics and Optics, School of Physics,
University of New South Wales, Sydney 2052, Australia}
 
\and
 
\author{R. L. Webster\altaffilmark{3}}
\affil{Astrophysics Group, School of Physics,
University of Melbourne, Victoria 3052, Australia}

\altaffiltext{1}{fmasci@physics.unimelb.edu.au}
\altaffiltext{2}{mjd@edwin.phys.unsw.edu.au} 
\altaffiltext{3}{rwebster@physics.unimelb.edu.au}

\begin{abstract}
The Parkes Half-Jansky Flat Spectrum Sample contains a large number of sources with
unusually red optical--to--near-infrared continua. If this is to be interpreted
as extinction by dust in the line-of-sight, then associated material might also
give rise to absorption in the soft X-ray regime. This hypothesis is tested using
broadband (0.1-2.4 keV) data from the {\it ROSAT} All-Sky Survey provided 
by Siebert et al.\ (1998).
Significant ($>3\sigma$ confidence level)
correlations between optical (and near-infrared)--to--soft X-ray continuum slope and
optical extinction are found in the data,
consistent with absorption by material with metallicity
and a range in gas-to-dust ratio as observed in the local ISM. 
Under this simple model, the soft X-rays are absorbed at a level consistent with 
the range of extinctions ($0< A_{V}< 6$ magnitudes) 
implied by the observed optical reddening.
Excess X-ray absorption by warm (ionised) gas, 
(ie. a `warm absorber') is not required by the data. 
\end{abstract}


\section{Introduction}

There have been numerous studies reporting the presence of soft
($\lesssim2$keV) X-ray absorption in excess of that expected from the
galaxy towards radio quasars.
However, such studies have found very little evidence for associated
optical reddening by dust (Elvis et al.\ 1994).  
Strong evidence for associated \ion{Mg}{2} and
soft X-ray absorption in a number of radio loud quasars has been
confirmed (eg. Mathur et al.\ 1994 and Mathur 1994), though evidence for associated
optical extinction in these sources is weak.
The lack of associated optical extinction may be due to a selection bias.
Since X-ray absorption estimates are
derived from spectra which require relatively large X-ray counts, such
studies may be biased against those sources with low counts due to
strong X-ray (and hence optical) absorption.
It is also possible that X-ray absorption in most sources 
is dominated by the presence of warm (ionised) absorbers
close to the primary continuum source (Pan, Stewart \& Pounds 1990).
These have little broadband effect in the optical bandpass but may strongly
absorb X-ray emission.
 
If strong optical extinction by dust is known a-priori however, then
associated absorption of soft X-rays is expected to be present at some
level.  Evidence for excess soft X-ray absorption in a number of
optically reddened radio quasars has been presented by 
Kollgaard et al.\ (1995). 
These authors claim that their results are
strongly model dependent, and are also consistent with explanations other
than absorption by associated gas and dust.  Their
sample size is also too small to draw any reasonable
conclusions.  Evidence for an association was recently reported
by Puchnarewicz et al.\ (1996) for a large sample of Seyfert 1s and
quasars.  These authors found a correlation between optical spectral slope 
and optical--to--soft X-ray continuum slope that was consistent with
absorption by dusty ``cold'' gas with column densities $>10^{21}{\rm
cm}^{-2}$ and approximately Galactic dust abundance.

In this paper we consider the Drinkwater et al.\ (1997)
sample of flat-spectrum radio sources (hereafter Parkes quasars), 
a subset of which are known to
be optically reddened (\cite{Webster95}, \cite{Francis98}).
Unfortunately, very little spectroscopic X-ray data exists for these quasars
to constrain absorption gas column densities.
Siebert et al.\ (1998) were able to measure power-law photon indices for 105
sources in the sample. They did not find any evidence for excess soft
X-ray absorption in the red sources, but they did find that (at low
redshifts) the red quasars had lower soft X-ray luminosities than the
blue quasars and concluded that the X-ray data support the importance
of dust in the reddest quasars.
 
We re-analyse the soft X-ray (0.1-2.4keV) {\it ROSAT} broadband measurements 
from Siebert et al.\ (1998) using a different approach.
We consider a subsample of 119 out of the 323 sources 
in the Drinkwater et al.\ sample, all of which have contemporaneous
$B_{j}$ and $K_{n}$ photometry, useful for estimates of spectral slopes. 
About half of these sources are detected in soft X-rays and upper limits
are available for the remainder.
Our analysis uses a simple gas-dust absorption model to explore various 
correlations involving optical (and near-IR)--to--soft X-ray continuum slope
and optical reddening. We then use the broadband soft X-ray data to search for
such correlations and determine whether
soft X-rays are absorbed at a level consistent with the observed optical reddening.  

\section{Quasar Sample and X-ray Data}
\label{alldata}

The Parkes Half-Jansky Flat Spectrum Sample contains 323 sources and is
described in detail by Drinkwater et al.\ (1997). 
The sample was initially selected from the Parkes 2.7 GHz Survey 
(Bolton, Savage \& Wright 1979 and references therein) according to the following
criteria:
2.7 GHz radio flux densities $f_{2.7{\rm GHz}}>0.5$ Jy, radio spectral indices
$\alpha_{2.7}^{5}<0.5$ (where $f_{\nu}\propto\nu^{-\alpha}$), Galactic latitude
$|b|>20^{\circ}$ and B1950 declinations: $-45^{\circ}<\delta<+10^{\circ}$.
By selecting flat-spectrum radio sources at high frequency, one is biased
towards core-dominated quasars, since lobe dominated quasars and radio galaxies
have steeper radio spectra and hence are likely to comprise the 
majority of detections 
in low frequency surveys.
On the basis of spectroscopic identification alone,
the present sample contains a much higher 
quasar fraction ($\gtrsim85\%$) than
any existing radio sample, with a broad and flat distribution in redshift to $z\sim4$.

For the purposes of this paper we have used sources with contemporaneous photometric
measurements in $B_{j}, V, R, I, J, H, K_{n}$, which are 
complete for a subsample of 119
`non-extended' sources.
These were obtained by P. Francis (Private communication; see Whiting et al.\ 1998)
in April, July and September 1997 using the ANU 40\arcsec~ and 2.3m telescopes.
These measurements provide reliable
estimates of broadband colours, likely to be unbiased 
with respect to uncertainties from
intrinsic variability.
RMS uncertainties based on noise statistics in these magnitudes are $<0.2$ mag.

Since we are primarily interested in the properties of
quasars, only spatially unresolved sources in the optical and near-infrared
have been considered.
All sources appearing extended in $B_{j}$ and $K_{n}$, or with optical spectra showing
features characteristic of those seen in 
normal nearby galaxies are excluded. We also ensured that the sources 
have broad emission lines in their optical spectra with 
velocity widths $>2000{\rm km}\,{\rm s}^{-1}$ (at FWHM), typical of normal QSOs.
Redshifts are available for all the sources and
span the range $0<z\leq3.9$, most of which are from
new spectroscopic observations (see Drinkwater et al.\ 1997).

X-ray data for all Parkes quasars has been provided by Siebert et al.\ (1998).
The fluxes are in the soft X-ray band 0.1-2.4 keV,
and most were determined from the {\it ROSAT} 
All-Sky Survey and (for 49 sources)
from archival pointed
PSPC observations. Of the 323 sources in the Drinkwater et al. sample, 163 were
detected in soft X-rays at the $3\sigma$ level. For the remaining 160 sources,
$2\sigma$ upper limits to the counts were determined. We have used 
the total broadband fluxes as computed by Siebert et al. from the count rates,
corrected for Galactic absorption only.
Where available, we also use their estimates of the photon energy indices $\Gamma$ 
(where $f_{\nu}\propto\nu^{1-\Gamma}$) determined from
explicit power-law fits to spectral data or from hardness ratio techniques
(see \S~\ref{slopes}).  
In total, for our subsample of 119 Parkes quasars with
contemporaneous photometry, 57 are detected in soft X-rays and upper limits
are known for the remaining 62.

\section{Optical Extinction and X-ray Absorption Modelling}
\label{model}

As claimed by Webster et al.\ (1995), if the large spread in
optical--to--near-IR colours of Parkes quasars is due to reddening by dust,
then it is expected that the reddest quasars may also be absorbed in soft X-rays.
We test this hypothesis by making simple predictions 
involving optical--to--soft X-ray ($\alpha_{BX}$) and
near-IR--to--soft X-ray ($\alpha_{KX}$) continuum slopes.
This section will briefly outline our assumptions and predictions using a simple
gas-dust absorption model.       

The degree of X-ray absorption by metal enriched gas primarily depends
on the total column density of gas in the line-of-sight. In the case of the galactic
ISM where the metal abundance is typically $\lesssim1\%$ relative to hydrogen by
mass (Grevesse \& Anders 1991), hydrogen and helium are responsible for almost
all of the absorption at energies $\lesssim2$keV.
We predict the amount of X-ray absorption expected for a given optical
dust extinction measure by assuming for simplicity, the range in
Galactic gas-to-dust ratios
derived empirically from
Ly-$\alpha$ absorption measurements in the Galaxy (Bohlin et al.\ 1978):
\begin{equation}
N({\rm HI+H_{2}})_{tot}\,\simeq\,(5.8\pm2.5)\times10^{21}\left(\frac{E_{B-V}}{\rm mag
}\right) 
\,\,{\rm cm}^{-2},
\label{ext_col}
\end{equation}
where $E_{B-V}$ is the extinction (colour excess) in $B-V$ colour.
This relation is
also consistent with empirical estimates of the dust-to-gas ratio in
the SMC and LMC by Bouchet et al.\ (1985) and Fitzpatrick (1985).

The effective X-ray optical depth, $\tau_{X}$, defined such that the 
change in flux due to absorption is $\exp(-\tau_{X})$, is given by:
\begin{equation}
\tau_{X}\,=\,\sigma_{E}N({\rm HI+H_{2}})_{tot}, 
\label{tauX}
\end{equation}
where $\sigma_{E}$ represents the effective absorption cross-section per H atom
at energy $E$ and $N({\rm HI+H_{2}})_{tot}$ is defined by equation (\ref{ext_col}).
We adopt cross-sections for X-ray absorption as derived by 
Morrison \& McCammon (1983) for a gas with galactic ISM metal abundances.
Typically, $\sigma_{\rm 1keV}\simeq2.42\times10^{-22}{\rm cm}^{2}$ 
and $\sigma_{\rm 2keV}\simeq4.30\times10^{-23}{\rm cm}^{2}$. 
Together with
the ratio of total to selective extinction $A_{V}/E_{B-V}\simeq3.05\pm0.15$ as given by
Whittet (1992), the optical depths at 1keV and 2keV from equation (\ref{tauX})
can be written respectively:
\begin{equation}
\tau_{\rm 1keV}\,\simeq\,(0.46\pm0.19)\,\left(\frac{A_{V}}{\rm mag}\right),
\label{tauAv}
\end{equation}
$$
\tau_{\rm 2keV}\,\simeq\,(0.08\pm0.04)\,\left(\frac{A_{V}}{\rm mag}\right).
$$
 
As a simple estimate, if a source at redshift $z=1$, (the median
redshift of the Parkes quasar sample) suffers an intrinsic
extinction $A_{V}=2$ mag (a typical mean value for Parkes quasars; Masci 1997), then
absorption by associated neutral gas would  reduce the {\it observed} 1keV flux
by about a factor $1/\exp(-\tau_{(1+z){\rm 1keV}})\sim1.2$. 
For comparison, assuming the generic
$1/\lambda$ dust law, an intrinsic extinction of $A_{V}=2$ mag will
result in a decrease of the {\it observed} optical flux by a factor $>50$.
Thus, we see that the decrease of observed optical flux by dust extinction
is generally larger than the corresponding decrease in soft X-ray flux by
associated gas. Even for low $z$ sources, the discrepancy in flux reductions
in these two regimes is about a
factor of 5, and increases rapidly with $z$ due to the energy dependence
of $\sigma_{E}$. This property will become
important when we examine the correlation involving optical--to--X-ray
continuum slopes in \S~\ref{apred}. 

\subsection{$\alpha_{BX}$ and $\alpha_{KX}$ versus Optical Extinction}
\label{apred}

Based on optical and X-ray spectral properties of Parkes quasars, 
both Webster et al. (1995) and Masci (1997) claimed that in most cases, 
the reddening
is likely to be due to dust {\it intrinsic} to the quasars.  Thus for
simplicity, the following analysis will assume all 
absorbing material to be located at the
redshift of the quasar.  We also assume that all predicted quantities discussed below
(ie. slopes and extinction measures) refer to the 
quasar {\it rest frame}.

If a source with some intrinsic (unabsorbed) optical--to--X-ray power-law slope, say
$\alpha_{BX_{i}}$ (where $f_{\nu}\propto\nu^{-\alpha}$) suffers
intrinsic absorption by dusty gas, then in its rest frame,
the resulting slope ($\alpha_{BX_{o}}$) can be written in terms of extinction
optical depths as follows:
\begin{equation}
\alpha_{BX_{o}}\,=\,\alpha_{BX_{i}} +
\frac{\tau_{X}-\tau_{B}}{\ln{(\nu_{X}/\nu_{B})}}.
\label{alphaBX}
\end{equation}
We shall also consider the near-infrared--to--X-ray continuum slope 
($\alpha_{KX_{o}}$), 
so that analogously,
\begin{equation}
\alpha_{KX_{o}}\,=\,\alpha_{KX_{i}} +
\frac{\tau_{X}-\tau_{K}}{\ln{(\nu_{X}/\nu_{K})}}.
\label{alphaKX}
\end{equation}
According to the available data, our anlysis will assume the $B_{j}$-bandpass 
($\lambda=475$nm) for the optical and the $K_{n}$-bandpass 
($\lambda=2.15\mu$m) for the near-infrared.
We shall consider the X-ray flux at 1keV, so that the  optical depth $\tau_{X}$ in 
equations (\ref{alphaBX}) and (\ref{alphaKX}) is defined in terms of the
optical extinction $A_{V}$ by equation (\ref{tauAv}).
Using the extinction coefficients $R_{\lambda}=A_{\lambda}/E_{B-V}$ from
Savage \& Mathis (1979) in the optical and Whittet (1988) in the near-infrared
for diffuse galactic dust, we find
\begin{equation}
\tau_{B}\simeq(1.223\pm0.015)A_{V}{\hspace{5mm}}{\rm and}{\hspace{5mm}}
\tau_{K}\simeq(0.084\pm0.030)A_{V}.
\label{tauBK}
\end{equation}
By combining the above relations, we can see from equations (\ref{alphaBX}) and
(\ref{alphaKX}) that for given 
intrinsic (unabsorbed) values $\alpha_{BX_{i}}$ and $\alpha_{KX_{i}}$,
absorption by dusty-gas will predict a specific correlation between optical
extinction ($A_{V}$) and the corresponding absorbed continuum slopes.

To make some illustrative predictions applicable to radio quasars, 
the intrinsic slopes
$\alpha_{BX_{i}}$ and $\alpha_{KX_{i}}$ need to be specified.
As a working measure, we assume for simplicity, these slopes to be
those found for {\it optically-selected} quasars. 
As argued by Masci (1997) and Francis et al.\ (1998),
this choice is based on the claim that optically-selected quasars
are expected to be strongly biased against significant absorption by dust. 
This indeed is consistent by a number of studies which find a relatively
small scatter in the optical--to--X-ray (1keV) flux ratio of optically-selected 
quasars (eg. Kriss \& Canizares 1985, Wilkes et al.\ 1994, La Franca et al.\ 1995).
The distributions in this flux ratio indicate a mean value 
$\langle\alpha_{BX_{i}}\rangle\sim1.3$ with dispersion
$\sigma\sim0.2$, and which does not significantly differ
between radio-loud and radio-quiet quasars (Wilkes \& Elvis 1987, Green et al.\ 1995).
This value is also consistent with that implied
by a composite SED
for radio loud (optically-selected) quasars 
derived by Elvis et al.\ (1994).  
Due to the absence of sufficient near-infrared data for optically selected samples,
we adopt the $K$-to-1keV continuum slope as indicated by this composite where
$\langle\alpha_{KX_{i}}\rangle\sim1$, and assume a scatter similar to that found for
$\alpha_{BX_{i}}$ in the studies above.
From these studies, our predictions will assume the following
ranges:
\begin{equation}
1.1\,<\,\alpha_{BX_{i}}\,<\,1.6,
\label{intslopes}
\end{equation}
$$
0.8\,<\,\alpha_{KX_{i}}\,<\,1.3.
$$

The radio-selected 
Parkes quasars for comparison show a dispersion in $\alpha_{BX}$ almost five times 
greater 
(see \S~\ref{rescomp}) than those of
optically-selected quasars (eq. [\ref{intslopes}]). 
Siebert et al. (1998) have shown that this is consistent with
dust-gas absorption, however in addition, the larger scatter could also 
include effects from
any of the following: an additional X-ray emission 
component over and above that of radio-quiet quasars with a wide
distribution of strengths and/or slopes, enhanced X-ray variability, or
a strong angle dependence of the observed X-ray emission due to beaming.
Although such effects can be important in some sources and contribute
to the scatter, our analysis 
here is purely concerned with the hypothesis 
that it is mostly due to variable amounts of absorption in the optical
and X-ray bands.   
 
Given the ranges in intrinsic slopes defined by equation (\ref{intslopes}) and
the ranges defined by the uncertainties in the extinction measures $\tau_{\rm 1keV}$
(eq. [\ref{tauAv}]) and $\tau_{B}$, $\tau_{K}$ (eq. [\ref{tauBK}]), we
show in Figure~\ref{pred} the expected range spanned by $\alpha_{BX_{o}}$ and
$\alpha_{KX_{o}}$ as a function of $A_{V}$.
There appears to be a distinct behaviour in each of these slopes with $A_{V}$,
with an anti-correlation in $\alpha_{BX_{o}}$ and correlation in $\alpha_{KX_{o}}$.
As discussed in \S~\ref{model} and as seen in equations (\ref{alphaBX}) and (\ref{alphaKX}),
this is due to the amount of absorption expected in $B_{j}$ and $K_{n}$ relative to
that at 1keV. From the above discussion, we have typically: 
$\tau_{B}/\tau_{X}\sim2.7$ and $\tau_{K}/\tau_{X}\sim0.2$, so that
the correlation involving $\alpha_{KX_{o}}$ is purely due to the fact that
absorption causes a greater
decrease in soft X-ray flux than that in the near-infrared.
It is important to note that the trends
of these correlations 
sensitively depend on the assumed dust-to-gas ratio
(eq. [\ref{ext_col}]) and metallicity 
(ie. the cross-section $\sigma_E$ in eq. [\ref{tauX}]),
which for simplicity were fixed to local ISM values. 
A variation in either of these quantities by at least a factor
of five is likely to change the sign of the correlation.

The predictions in Figure~\ref{pred} are represented by the 
following linear relations:
\begin{equation}
\alpha_{BX}\,=\,(1.3_{-0.2}^{+0.3}) - (0.13_{-0.02}^{+0.03})\times A_{V}
\label{linrel}
\end{equation}
$$
\alpha_{KX}\,=\,(1.0_{-0.2}^{+0.3}) + (0.05_{-0.03}^{+0.04})\times A_{V}.
$$
Given the uncertainties, the slope of the $\alpha_{KX}$ versus
$A_{V}$ relation is insignificant, however for typical galactic ISM conditions, 
the crucial feature in Figure~\ref{pred} is the opposite trend predicted in 
each of these slopes
in the presence of absorption
by dusty gas. This will provide a 
powerful diagnostic for testing the
dust reddening hypothesis for Parkes quasars.

\placefigure{pred}

\section{Comparison with Data}

This section will compare the predictions of Figure~\ref{pred} with
estimates of the slopes involving the soft (0.1-2.4 keV) X-ray bandpass
($\alpha_{BX}$ and $\alpha_{KX}$) derived
from the available {\it ROSAT} data, and extinctions ($A_{V}$)
derived from the observed optical reddening.
To facilitate a direct comparison with the predictions of Figure~\ref{pred}
however, we first 
transform these quantities into the source rest frame.

\subsection{Spectral Slopes and Transforming to the Rest Frame}
\label{slopes}

Due to the frequency dependence of optical extinction and 
X-ray absorption, a similar dependence of these
quantities on the redshift of the absorbing material in
an {\it observer's} frame is expected. 
This implies that a plot similar to Figure~\ref{pred} which uses
{\it observed} quantities will also include a hidden and complicated 
dependence of
the absorption on redshift. The added
effects of changing spectral shape with source redshift
(ie. K-correction effects) will also introduce ambiguities.
Thus, assuming the absorbing material is
intrinsic to the quasars, we will transform all observed quantities 
to the source rest frame. This will enable
a direct and unambiguous comparison with the predictions of Figure~\ref{pred}.

We have used the total 0.1-2.4 keV X-ray fluxes 
as computed from the {\it ROSAT} count
rates by Siebert et al.\ (1998), which were corrected for
Galactic absorption only.
The {\it observed}
spectral indices (ie. $\alpha_{BX}({\rm obs})\propto\log{[f_{x}/f_{B}]}$),
were computed by first deriving monochromatic X-ray fluxes $f_{x}$ at 1keV. 
These were determined from the total broadband
fluxes and available upper limits
assuming a power-law continuum ($f_{\nu}\propto\nu^{-\alpha_{X}}$)
between 0.1-2.4 keV.
Where available, we have used the photon indices $\Gamma$ 
(where $\alpha_{X}\equiv\Gamma - 1$)
determined from explicit
power-law fits to spectral data and hardness ratio techniques by Siebert et al.\ (1998).
These were available for 71 of the 119 sources. 
When an individual photon index was not available, the average photon
index for radio-loud quasars, $\alpha_{X}=1$ (eg. Schartel et al.\ 1996) was used.

The rest frame spectral indices, for example $\alpha_{BX}({\rm rest})$, were determined
by applying a simple K-correction which assumes a power-law in each of the optical
and soft X-ray bands. If $\alpha_{B}$ and $\alpha_{X}$ are respectively the
optical and X-ray power-law slope, $z$ the source redshift and 
$\alpha_{BX}({\rm obs})$ the {\it observed} optical--to--X-ray 
slope, then the corresponding value
in the rest frame is given by:
\begin{equation}
\alpha_{BX}({\rm
rest})\,=\,\frac{(\alpha_{B}-\alpha_{X})\log{(1+z)}}{\log{(\nu_{X}/\nu_{B})}}\,+\,
\alpha_{BX}({\rm obs}).
\label{Kcorr}
\end{equation}
X-ray slopes $\alpha_{X}$ are taken from Siebert et al.\ (1998) as discussed above,
and optical slopes $\alpha_{B}$ for each source 
were determined from the contemporaneous
photometry, measured between the $B_{j}$ and $I$ ($\simeq0.9\mu$m) passbands.
In our determination of the near-IR--to--X-ray slope $\alpha_{KX}({\rm rest})$,
the required near-IR slopes $\alpha_{K}$ were determined between the
$B_{j}$ and $K_{n}$ passbands.

Rest frame optical extinctions $A_{V}$ in each source were derived from the
observed optical--to--near-IR reddening as defined by the 
contemporaneous colours $B_{j}-I$ and $B_{j}-K_{n}$ and assuming
intrinsic (unabsorbed) colours as measured in optically-selected quasars.
As discussed in \S~\ref{apred}, optically-selected quasars are expected to be
strongly biased against significant absorption by dust. This claim is consistent
with their
relatively small
scatter in colours (eg. Francis 1996), 
which lie
predominately on the blue tail of the Parkes-quasar colour distribution.
From quasi-simultaneous optical/near-IR photometry by Francis (1996) of 
a subset of 37 quasars drawn from the optically-selected LBQS sample 
of Hewett et al.\ (1995), we find the following mean values and dispersions in
intrinsic colours:
\begin{equation}
(B-K)_{i}=2.3\pm0.5{\hspace{5mm}}{\rm and}{\hspace{5mm}}(B-I)_{i}=0.9\pm0.4.
\label{intC}
\end{equation}
We assume these values represent the intrinsic (unabsorbed) colours
of Parkes-quasars.
For our redshift range of interest: $0\leq z\leq3$,
we also find that for these optically-selected quasars, the colours 
show no significant dependence on redshift. 
We therefore assume the intrinsic colours (eq. [\ref{intC}])
to be independent of redshift
in an {\it observer's frame}.

Given a general dust extinction curve defined by 
$\xi(\lambda)\equiv A_{\lambda}/A_{B}$,
the {\it rest frame} optical extinction can be written in terms of
an observed colour, say $(B-K)_{o}$, and corresponding intrinsic colour $(B-K)_{i}$
as follows:
\begin{equation}
A_{V}\,=\,\left(\frac{\xi(\lambda_{V})}{\xi(\lambda_{B}/1+z)-
\xi(\lambda_{K}/1+z)}\right)
\left[(B-K)_{o}-(B-K)_{i}\right],
\label{Avc}
\end{equation}
where $z$ is the source redshift. We have used the analytical fit for $\xi(\lambda)$
as derived by Pei (1992) for diffuse galactic dust in the range 
$500{\rm\AA}\lesssim\lambda\lesssim 25\mu{\rm m}$.

The {\it rest frame} extinction $A_{V}$ for each 
source was estimated by computing the
average of the two extinction values obtained independently from the two observed 
colours
$B_{j}-I$ and $B_{j}-K_{n}$.
As shown in Figure \ref{av}, there is a tight correlation between these
estimates for $A_{V}$, suggesting each reddening indicator may equally
provide a measure of the extinction. From the scatter about the line of equality, the
values are consistent to within $\pm0.5$ mag.
This correlation suggests that the $B-K$ and $B-I$ colours of Parkes-quasars
vary with each other in such a way implying a characteristic 
spectral curvature from $B$ to $K$ that is
consistent with reddening by dust. This is unlikely to arise from
intrinsic relationships between these bandpasses in the source emission, such as
a correlation between the $I$ and $K$ bands. Appropriate fine tuning would
be required to reproduce the result in Figure \ref{av}. 

\placefigure{av}

\subsection{Results and Model Comparisons}
\label{rescomp}

Using the above formalism to convert the observed quantities into the
source rest frame, Figure \ref{data} shows the rest frame spectral indices
$\alpha_{BX}$ and $\alpha_{KX}$ as a function of the optical extinctions.
The triangles represent $2\sigma$ lower limits on these indices from the
X-ray non-detections. In the lower left corners of each figure, we
show two `conservatively' calculated error bars 
for both the $A_{V}$ and spectral indices. From the dispersions in our assumed intrinsic
colours (eq. [\ref{intC}]) and scatter in Figure~\ref{av}, we have typically:
$\sigma(A_{V})\simeq0.5$ mag. The spectral index error bar assumes a maximum error in
the X-ray flux of 25\% 
(ie. $\sigma(\log{f_{x}})=0.25$; see Siebert et al.\ 1998) 
and an uncertainty in both the $B_{j}$ and
$K_{n}$ bands of 0.2 mag (see \S~\ref{alldata}). 
Furthermore, following Siebert et al.\ (1998)
we also included an uncertainty for possible variability in the non-simultaneous
X-ray and $B_{j}$ (or $K_{n}$) measurements. This assumes a variability of
$\sigma(B)=\sigma(K)=0.3$ mag, typical for one of our 
quasars with a 20 year rest-frame timescale
between the optical and X-ray measurements (Hook et al.\ 1994). Combining these
uncertainties, we have $\sigma(\alpha_{BX})=\sigma(\alpha_{KX})\simeq0.15$.

The optical--to--X-ray slopes (Fig.~\ref{data}a) and
near-infrared--to--X-ray slopes (Fig.~\ref{data}b) appear to be somewhat
anti-correlated and
correlated with $A_{V}$ respectively.  
We have formally computed the probabilities that the observed correlations are
real by taking into account all lower limits on the spectral slopes
and using the techniques of survival analysis.
The correlation and regression analyses were performed using the ASURV package
(Version 1.3; 
La Valley et al.\ 1992), which is particularly designed for censored data,
implementing the methods presented in Isobe et al.\ (1986).
For the correlation analysis we applied the generalised Kendall's tau test
and for the regression analysis we used the parametric EM algorithm.
For $\alpha_{BX}$ vs. $A_{V}$ (Fig.~\ref{data}a) we find that the probability
for {\it no} correlation is $P=0.0006$, i.e. the hypothesis that these quantities
are uncorrelated is rejected at the 99.94\% confidence level.
For $\alpha_{KX}$ vs. $A_{V}$ (Fig.~\ref{data}b), we find $P<0.0001$, the
correlation is significant at the $>99.99\%$ confidence level.
The solid lines in Figure~\ref{data} are our best regression fits to the data. Using
ASURV, we have the following best regression line fits:
\begin{equation}
\alpha_{BX}\,=\,(1.41\pm0.02) - (0.08\pm0.02)\times A_{V} 
\label{reg}
\end{equation}
$$
\alpha_{KX}\,=\,(1.18\pm0.03) + (0.10\pm0.02)\times A_{V}.
$$

The observed anti-correlation and correlation involving $\alpha_{BX}$ and
$\alpha_{KX}$ respectively
appear broadly 
consistent with the predictions of our simple model in Figure~\ref{pred}.
The predicted ranges in these slopes as a function
of the optical extinction (taking account uncertainties in dust-to-gas ratios
and intrinsic slopes; see \S~\ref{model}) are represented by the regions within 
the dashed curves in Figure~\ref{data}.
For $A_{V}\simeq0$, we see there is very good agreement
between the observed slopes of Parkes quasars and our
assumed range in intrinsic (unabsorbed) values (eq. [\ref{intslopes}]) from
studies of optically-selected quasars.
The optical (and near-IR)--to--soft X-ray continuum slopes of the bluest
Parkes quasars therefore show relatively small scatter similar to those of
optically-selected quasars, strengthening the claim that
such sources are unbiased with respect to reddening by line-of-sight dust.
The increased scatter in these slopes when all Parkes quasars are considered can
then be attributed to dust extinction. In particular, the range in extinctions:
$0< A_{V}<6$ mag, are consistent (within our 
conservative errors) with the maximum extinction $A_{B}\sim4$ mag derived by 
Siebert et al.\ (1998) from the dispersion in optical--to--X-ray flux ratio alone.

Our results are also consistent with studies of the optical--to--soft X-ray continua
of a large sample of medium to moderately-hard X-ray selected AGN by 
Puchnarewicz et al.\ (1996). Their study suggests 
moderate absorption by dusty gas with 
approximately galactic dust-to-gas ratio and columns
$N_{\rm H}<5\times10^{21}{\rm cm}^{-2}$.
Using the lower limit $N_{\rm H}/E_{B-V}\simeq3.3\times10^{21}{\rm cm}^{-2}
{\rm mag}^{-1}$ 
from equation (\ref{ext_col}) and a galactic extinction curve, this
corresponds to $A_{B}\lesssim6$ mag, entirely consistent with the range found 
in our sample of radio-selected quasars. 

\placefigure{data}

To summarise, we have found significant correlations between
optical (and near-IR)--to--soft X-ray continuum slope and optical extinction
that are consistent with the predictions of a simple dust model. 
The results of Figure~\ref{data} imply that if the observed scatter in 
optical slopes 
of Parkes-quasars is due to extinction by dust in their environs, then
soft X-rays are absorbed at a level fully consistent with this hypothesis.
    
\section{Discussion}

The anti-correlation involving $\alpha_{BX}$ and optical extinction 
(Fig.~\ref{data}a) can be easily explained by gas-dust absorption whereby
optical flux is relatively more absorbed than that at 1keV.
This anti-correlation is qualitatively similar to that
claimed by McDowell et al.\ (1989) for a sample of quasars which appeared
to show a weak ``big blue bump'' feature relative to their near-infrared
and soft X-ray emission. These authors however interpreted the trend as due to
an intrinsically varying blue-bump spectrum and not extinction by dust.
A changing blue luminosity may thus mimic a variation in optical extinction. 
It is therefore possible that the observed correlation in 
Figure~\ref{data}a is spurious and
due to a secondary trend, ie. from a dependence of both $\alpha_{BX}$ and
$A_{V}$ (or optical slope) on $B$-band luminosity.
A direct investigation of the relationship between these quantities
however shows that such an effect is unlikely to explain
the observed trend.
Also, since the optical-UV flux in the blue-bump
component also provides a source of photoionizing
flux for emission line regions, this model also predicts to first order a
correlation between  
the equivalent widths of emission lines and optical-UV continuum
slope. Our analysis of the emission line equivalent widths for Parkes quasars
(Masci 1997, Francis et al.\ 1998) however strongly argues against this possibility.
An intrinsically varying optical/UV continuum will also require the X-ray continuum
to vary simultaneously to satisfy the correlation involving $\alpha_{KX}$ in
Figure~\ref{data}b. This however would be inconsistent with
the trend in Figure~\ref{data}a. 
Such a mechanism therefore requires physically unmotivated
fine tuning of
the near-infrared, optical and soft X-ray emission to explain the
observed correlations. 

The correlation involving $\alpha_{KX}$
(Fig.~\ref{data}b) indicates that all of the optical--to--1keV continuum flux
appears absorbed relative to that in the near-infrared. It is encouraging to
find that such a correlation is qualitatively similar to that
claimed by Ledden \& O'Dell (1983)
between radio--to--soft X-ray and radio--to--optical slope for several
optically reddened radio quasars. They concluded that absorption by 
associated gas and dust was the most likely explanation. Their statistics
however were too low from which to draw any firm conclusion.

The indirect confirmation for soft X-ray absorption from
Figure~\ref{data} implies that the soft 
X-ray, optical and possibly also the near-IR continuum emission must traverse the
same dust component.
As suggested by previous studies to explain the difference in soft X-ray
spectral properties of radio-loud and radio-quiet quasars,
a significant fraction of the
soft X-ray flux in radio-quasars 
is believed to arise from synchrotron self-Compton (SSC) emission
associated with the radio jet emission (eg. Wilkes \& Elvis, 1987,
Ciliegi et al.\ 1995). 
According to standard unified models for AGN,
the radio jet emission may extend to distances $\gtrsim10$kpc from the
central engine.
Thus, if the soft X-rays are mostly SSC in origin, then the results of
Figure~\ref{data} require the
absorbing medium to extend on a scale similar to that of the radio emission.
If the absorbing medium however were situated close to the
central AGN so that the SSC X-rays suffered minimal
absorption (with effectively $\tau_{X}\approx0$) then the predictions
of Figure~\ref{data} will change drastically:
a much steeper anti-correlation in
Figure~\ref{data}a and the opposite behaviour (ie. anti-correlation) in
Figure~\ref{data}b would be expected.
More direct studies of X-ray absorption, preferably via high quality spectral
observations are required to further explore this issue. 

\section{Conclusions}

We have
tested the dust-reddening hypothesis to explain 
the relatively large dispersion in 
optical--to--near-infrared colours of quasars in the Parkes sample of
flat-spectrum radio sources by searching for associated absorption by gas at 
soft X-ray energies. 
We have used broadband (0.1-2.4 keV) 
{\it ROSAT} All-Sky Survey data and pointed PSPC observations
provided by Siebert et al.\ (1998), and contemporaneous optical--near-infrared
photometry available for 119 of the 323 sources in this sample.

A soft X-ray absorption signature was searched for indirectly by exploring  the
optical (and near-infrared)--to--X-ray continuum properties as a function
of optical extinction, where specific strong correlations
are expected under a simple gas-dust absorption model.

Our main conclusion is that significant correlations (at $>3\sigma$ confidence)
are observed between rest frame optical (and near-infrared)--to--1keV 
continuum slope and
optical extinction that are consistent with
the predictions 
of a simple model. This model assumes 
the range in dust-to-gas ratios and metal abundances
derived empirically from the galactic ISM, intrinsic (unabsorbed) slopes
as observed in optically-selected quasars, and as suggested by previous studies, that
all absorption is intrinsic to the quasars.
Under these assumptions, 
we conclude that soft X-rays are absorbed at a 
level consistent with 
the range of extinctions $0< A_{V}< 6$ mag observed.

The dust associated X-ray absorption is therefore consistent with physical
conditions (eg. gas ionization state)
found in the diffuse local ISM. 
No warm (ionised) absorption is required, since the data does not 
indicate excess X-ray absorption relative to that in the optical compared
to the galactic model predictions. 
This is contrary to numerous previous studies
of the soft X-ray properties of AGN.
High quality X-ray spectra to
detect heavy metal absorption edges
however are required to place stronger constraints on the physical properties.
These data
will also be necessary to provide `direct' estimates of the amount of X-ray absorption
and for constraining the geometry of absorbing/emitting regions. 
Due to the faintness of many of the reddest sources, spectral data is currently unavailable, 
and requires the high signal-to-noise and
resolution capabilities of future X-ray missions like {\it AXAF} and {\it XMM}.
    
\acknowledgments

We are grateful to Siebert et al. for making available the {\it ROSAT} data
(currently in press) and to Paul Francis for providing
the unpublished contemporaneous photometry (to appear in Whiting et al.\ 1998;
in preparation).

\clearpage

\figcaption[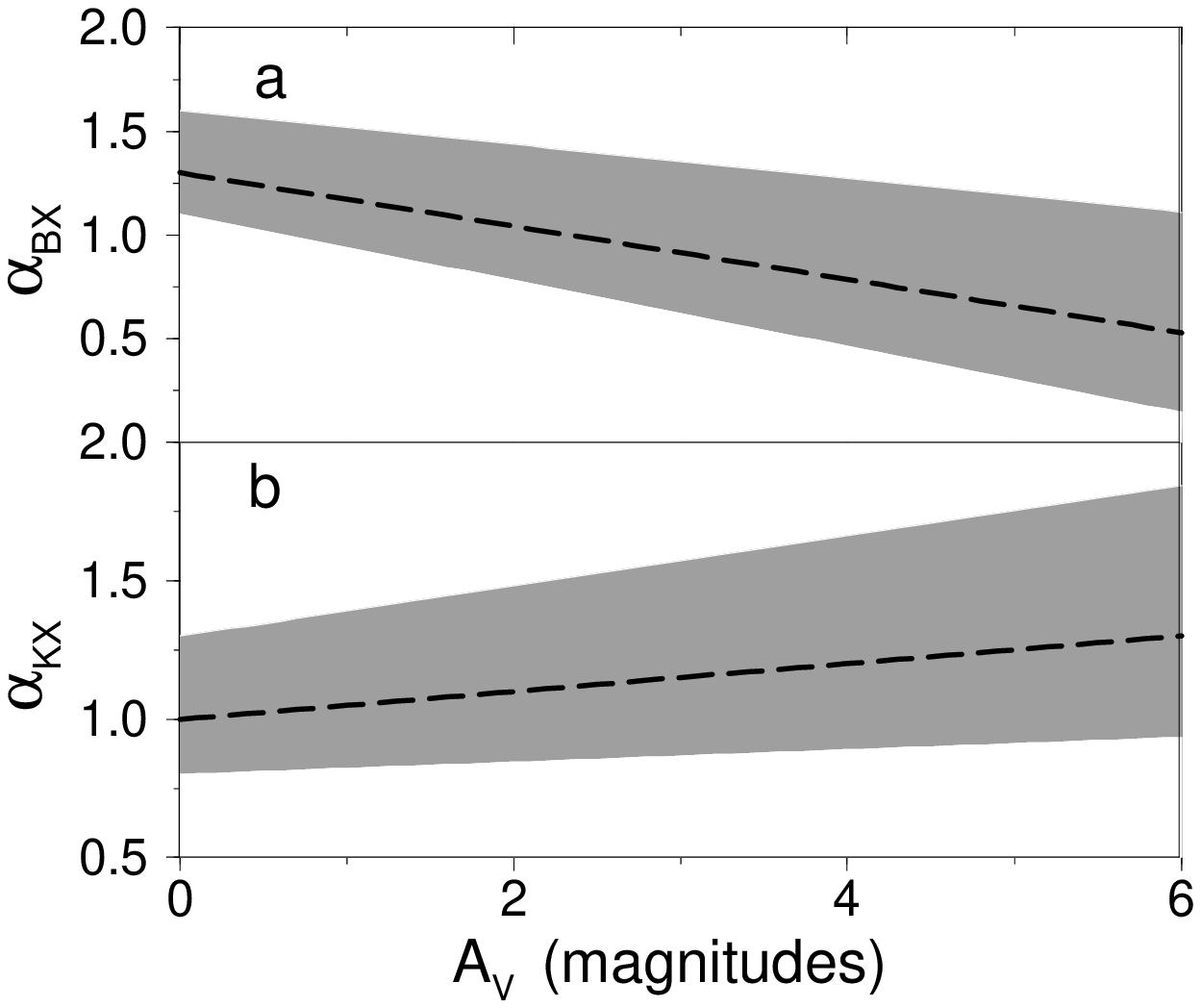]{(a) Rest frame $B_{j}$--to--1keV power-law continuum slope 
($f_{\nu}\propto\nu^{-\alpha}$), and
(b) $K_{n}$--to--1keV continuum slope as a function of optical extinction
$A_{V}$ predicted assuming the empirical ranges in 
galactic dust-to-gas ratio, extinction measures
and intrinsic slopes as discussed in \S~\ref{model}. Dashed lines represent mean
predictions.
\label{pred}}

\figcaption[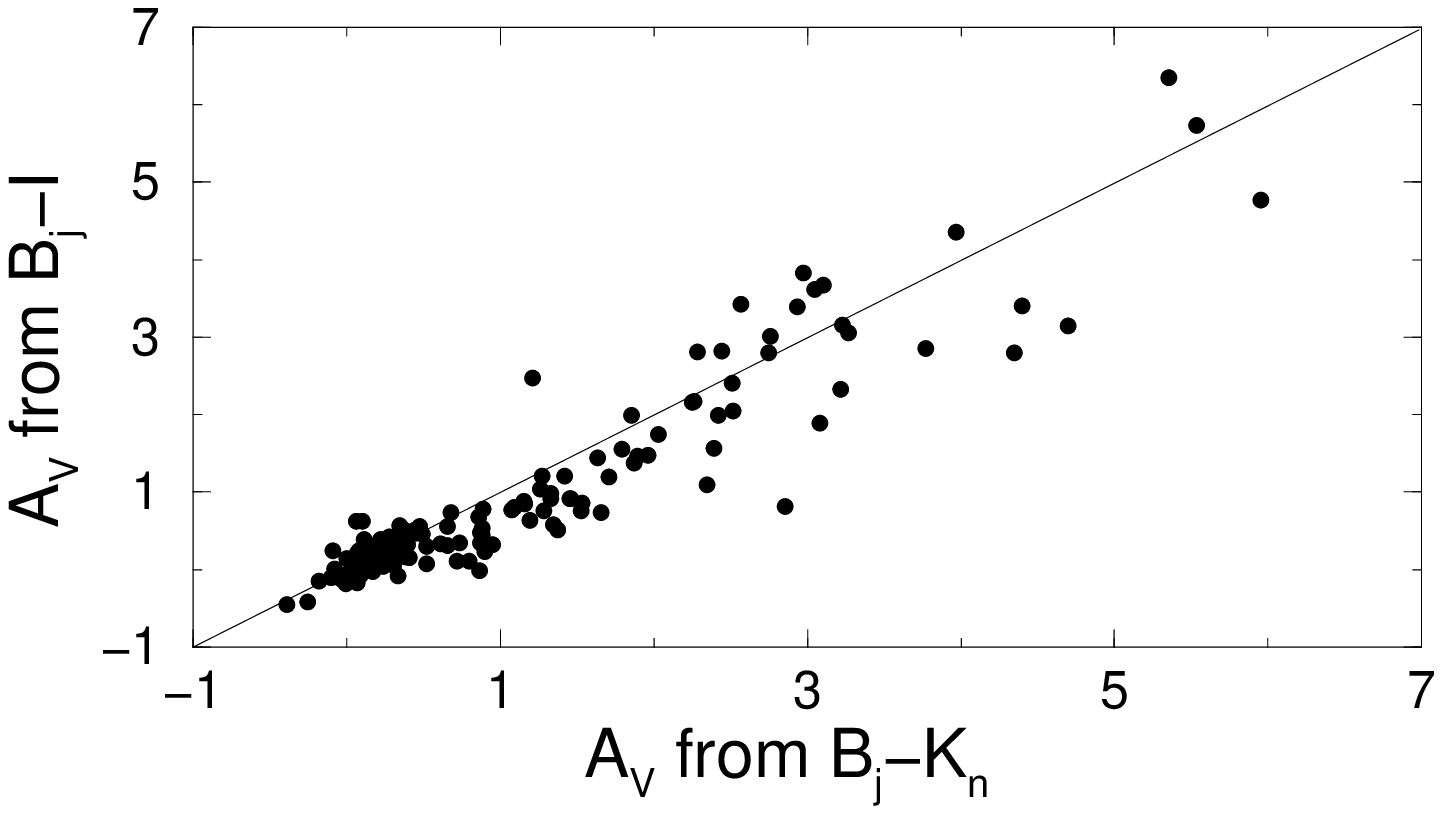]{Rest frame extinctions as determined from the observed
$B_{j}-I$ colours versus
those determined from $B_{j}-K_{n}$. The diagonal line is the line of equality.
\label{av}}

\figcaption[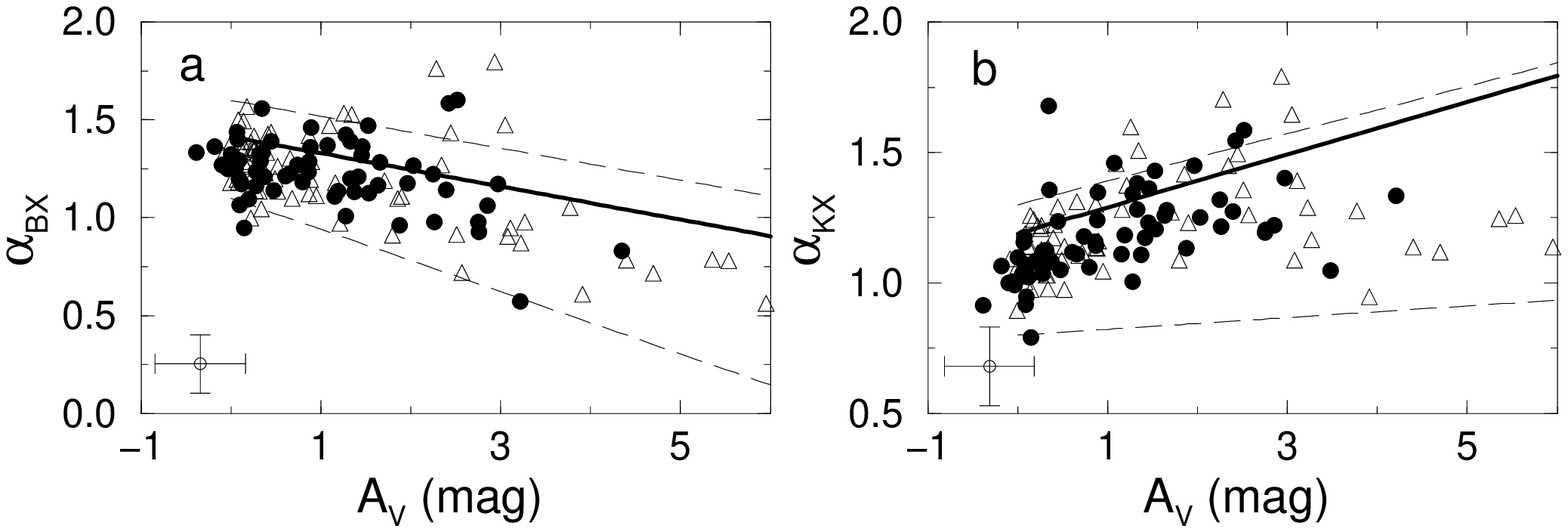]{(a) Rest frame $B_{j}$--to--1keV continuum slope
($f_{\nu}\propto\nu^{-\alpha}$), and (b) $K_{n}$--to--1keV continuum slope
as a function of rest frame optical extinction for Parkes quasars. Triangles
represent $2\sigma$ lower limits from the X-ray non-detections. Errors in both $A_{V}$
and slopes
are represented by conservatively calculated error bars
shown in the lower left, as discussed in \S~\ref{rescomp}.
Dashed lines represent the extremities of the range predicted by our simple model
(see Fig.~\ref{pred}). Solid lines are best line regression fits.
\label{data}}

\end{document}